\documentclass[12pt,aps,prb,preprint]{revtex4}
\usepackage{amssymb,amsfonts}
\usepackage[pctex32]{graphics}
\usepackage{graphics}
\newtheorem{theorem}{Theorem}[section]

\newtheorem{comment}{Comment}[section]
\newtheorem{definition}{Definition}[section]
\newenvironment{proof}{\smallskip\noindent{\textit{Proof $\;$}}}{$\Box$}
\begin{document}

\title{Predictive approach to some quantum paradoxes}
\author{Henryk Gzyl}
\affiliation{ IESA, Caracas, Venezuela}
\email{henryk.gzyl@iesa.edu.ve}
\date{}   
\begin{abstract}In classical probability theory, the best predictor of a future observation of a random variable $X,$ is its expected value $E_P[X]$ when no other information is available When information consisting in the observation of another random variable $Y$ is available, then the best predictor of $X$ is another random variable $E_P[X|Y].$ It is the purpose of this note to explore the analogue of this in the case of quantum mechanics. We shall see that exactly as in classical prediction theory, when the result of an observation is taken into account by means of a non-commutative conditional expectation, some of the usual paradoxes cease to be such. 
\end{abstract}
\maketitle
\section{Introduction and preliminaries}
In this section we recall the difference between measurement and prediction, and we describe what does classical prediction consist of, and finally we recall the basics about non-commutative conditional expectation necessary to present a quantum analogue to classical prediction theory.
\subsection{Prediction versus measurement}
Non differentiating between these concepts may be a source of confusion. Consider the following examples. Suppose you own some stock and want to decide whether to sell today or wait until tomorrow, or you want to answer one of the following the questions: What is the outcome of the toss of a fair die? What will the position and velocity of a particle at some future time be, if it moves under the action of a given force and we know its position and velocity now?

One possibility is to wait till tomorrow, and check the financial pages to find the value of your stock, or to toss the die and observe the outcome, or to measure the position at the specified time. That is, you may answer your question by means of a measurement.

But if you want to decide on a future course of action, you have to proceed differently. In the case of the particle, you may solve Newton's equations of motion and from the information  available now, make your prediction. In the other two cases, things are a bit more complicated: the relationship between the today's data to tomorrow's data is of a probabilistic nature. One thing you may do is to specify the probabilities of the possible results and leave it at that.  Or you may make a prediction. This consists of three steps: choosing a random variable (that is usually specified in the question to be answered). Choose a predictor, compute it and specify a measure of error in your prediction. 

The simplest predictor that is usually used used is the mean value: If $X$ denotes the random variable, the simplest predictor of the next outcome of $X$ is $E[X],$ the expected value of $X,$ and the simplest measure of errors are quantities of the type $P(|X - E[X]| > \epsilon)$ or $E[(X - E[X])^2].$ The quantity $\epsilon$ is the error with which the experimenter or decision maker feels confident with. It is important to keep in mind that a prediction is something made with ``pencil and paper,'' that is it does not involve measuring, regardless of the fact that is uses statistical data. It is akin to guessing the result of the next measurement. To beat it further, notice that the expected value does not need to be one of the possible results of a measurement, like the example of the die shows. 

Keeping this distinction between prediction and measurement is important when interpreting quantities like $<\psi,A\psi>.$ For example, if $A$ is an operator describing an observable, then for a given $\psi$, $(A - <\psi,A\psi>)^2$ is another operator describing an observable. The simplest predictor of the next measurement of that observable is $\sigma^2(A) = <\psi,(A  - <\psi,A\psi>)^2\psi>,$ which should be interpreted as the (expected) measurement error of $A$ in the state $\psi.$ 

Notice as well that, if $A,B,C$ are such that $[A,B] = iC,$ then it is a well known result that $\sigma(A)\sigma(B)\; \geq \; \frac{1}{2} |<\psi,C\psi>|,$ which is to be interpreted as: for any state $\psi$ the {\it predicted errors} in the measurement of $A,B,C$ are such that the inequality is satisfied, or there exists no $\psi$ for which the inequality is violated. But keep in mind that is an inequality about standard deviations or predicted measurement errors, not an inequality about results of actual measurements. Thus keeping in mind the distinction between measurement and prediction is essential.

\subsection{A remainder about prediction in classical science}

In classical mechanics in particular and in many mathematical models, the issue of prediction appears in essentially two different ways. On one hand we have systems, like in classical mechanics, whose dynamics, be it regular or irregular is deterministic. On the other hand we have systems, whose dynamics may be deterministic, but what it determines are transition probabilities, that is, if you know the probabilities of occurrence of the different states at $t=0$ and the ingredients of the Chapman-Kolmogorov equation, you may determine exactly the occupation probabilities at any later time.

For regular deterministic systems life is easy in principle, for statistical analysis is usually necessary only to deal with indeterminacy in initial data. When a system is deterministic but irregular, only short term prediction is feasible as for regular systems. For long term prediction we can recur to statistical prediction in the event the system is chaotic.

Prediction theory is a pretty much developed theory for classical (as opposed to quantum) stochastic systems. When dealing with systems for which there are no dynamics, like predicting the winner of an election, we recur to probabilistic modeling combined with statistical analysis of available data. It is here (and for classical stochastic systems) where the notion of conditional expectation has proved to be a key idea. Random variables are modeled as  measurable functions on a ``probability space"  $(\Omega, \mathcal{F}, \mathbb{P})$, where the ``sample space'' $\Omega$ is just a set, the questions we convene to ask about the system (namely the available information about the system) are the elements of a $\sigma$-algebra $\mathcal{F}$ of subsets of $\Omega$, and $\mathbb{P}$ is a measure on $\mathcal{F}$, assigning total mass 1 to $\Omega$. Given an integrable random variable $X$, without any further information, the best prediction that we can make about a future observation of that variable is $E\mathbb{P}[X] = \int Xd\mathbb{P},$ the expected value of $X$ with respect to $\mathbb{P}.$

Also, when $X$ is integrable, and we observe another random variable $Y$, we can ask, what is the best prediction about $X$ given that $Y$ has been observed. This best predictor happens to be another random variable, which in all cases of practical interest is a function of $Y$, that is denoted by $E[X|Y]$ and the following property that explains why it is the ``best predictor'': $E[X|Y]$ realizes $\inf E[(X - \phi(Y))^2]$ over all measurable, bounded functions $\phi.$ The formal definition of $E[X|Y]$  is that it is the unique function of $Y$ such that $E[Xg(Y)] = E[E[X|Y]g(Y)]$ for any bounded measurable function $g.$

That this predictor is a random variable, which is a function of $Y$, means that, for example, if $Y$ is discrete and takes values $\{y_1,y_2,...\},$ then,  when $Y$ is observed to assume the value $y_j$, then the value of $E\mathbb{P}[X|Y]$ is $E\mathbb{P}[X|Y = y_j] = \int X(\omega)d\mathbb{P}(\omega|Y = y_j)$ with probability $\mathbb{P}(Y = y_j).$ For the sake of comparison with the quantum case, in this case $E\mathbb{P}[X|Y]$ can be represented as
\begin{equation}\label{discreterepresentation}
E\mathbb{P}[X|Y] = \sum_{j}E\mathbb{P}[X|Y = y_j]I_{\{Y = y_j\}},
\end{equation}
\noindent where for any event $A,$ the indicator function $I_A$ is a dichotomic random variable taking  values $1$ or $0$ according to whether $\omega \in A$ or not. 

The properties of $E[X|Y]$ can be read in almost any probability book, and their use to analyze the classical analogues of some of the standard quantum paradoxes is carried out in Gzyl (2004), where the basic measure theoretic concepts are recalled.

A very simple example reminiscent of some quantum paradoxes, goes as follows. Consider  two binary random variables $X$ and $Y$ taking values $\pm 1$, and note that
$$P(X = -1 \,|\,X + Y = 0; Y =1) = 1.$$
The reader should supply the proof. The interpretation is obvious: Given that we are in state of total spin = 0 (i.e., given the event $X + Y = 0$ occurs), and that $Y = 1$, then our prediction is that $X = -1$ occurs with probability equal to $1$. We do not have to measure it, nor there is anything propagated between $X$ and $Y$. For comparison with the results for the non-commutative case, note that
in general $P(A|B,C) = P_C(A|B)$ where $P_C$ is the original probability conditioned upon the occurrence of $C,$ i.e., for any event $A$ we have $P_C(A) = P(A|C).$  In our example $C = \{X + Y = 0\},$ and also $P_C$ is carried by $C$ in the sense that $P_C(A) = 0$ whenever $P(A\cap C) = 0.$ Having said this, we add that another way to obtain the predicted value $E\mathbb{P}[X|Y=y_j]$ of $X$ when $Y = y_j$ is observed, is to compute 
$$E_\mathbb{P_j}[X] = E_\mathbb{P_j}[E_\mathbb{P}[X|Y]]$$ 
\noindent where $P_j$ is defined to be the conditional probability $\mathbb{P}(\bullet| Y = y_j).$ This has a quantum counterpart as we shall mention below.

\subsection{Conditional expectations in quantum mechanics}
Since the subject is rather technical, it is not surprising that it does not show up in (advanced) introductory books like the one by Peres (1993), nor at serious divulgative books like Accardi's (1997), Selleri's (1990) or Ghirardi's (2005); but it is not even mentioned in a nice advanced book like Landsman's (1998). We should also remark that two classics books on quantum estimation, like the volumes by Helstrom (1976) or Holevo (1982), do not even mention the possibility of using non-commutative conditional probabilities to develop the quantum analogue of classical prediction. The theme is not considered either in the monograph on quantum measurement theory by Bush et al (1991). The material below is taken form Gudder and Marchand's (1972), where references to the basic literature can be seen. But see also Gudder's (1979) where the comparison with classical probability theory is examined.

Our setup will be standard: A separable Hilbert space $\mathcal{H}$ is chosen to describe a particular system, we shall consider the von Neumann algebra $\mathbb{A}$ of all bounded operators on $\mathcal{H}$ and $\mathcal{P}_\mathbb{A}$ will denote the class of all self-adjoint projections on $\mathbb{A}.$ Two related concepts are contained in the following definitions. A measure is a mapping $w: \mathcal{P}_\mathbb{A} \rightarrow [0,\infty)$ such that (i) $w(0) = 0,$ and (ii) $w(\sum A_j) = \sum w(A_j)$ for any countable collection of mutually orthogonal $\{A_j\}$ in $\mathcal{P}_\mathbb{A}.$  A linear functional $w:\mathbb{A} \rightarrow \mathbb{C}$ is an integral if it satisfies (i) and (ii) above and also (iii) $w(A) \geq 0$ if $A$ is a positive element in $\mathbb{A}$.  When $w(I) = 1$ $w$ is called a state.

Let $w(A) = tr(WA)$, where $W$ is a self-adjoint, positive operator such that $tr(W) = 1.$ The operator $W$ is called the density operator of $w$.
Let us now recall
\begin{definition} With the notations introduced above, let $\mathbb{B} \subset \mathbb{A}$ be a sub-algebra of $\mathbb{A}.$  The $\mathbb{B}$-expectation of $A \in \mathbb{A}$ is an operator $E_w
[A|\mathbb{B}]$ satisfying $w(BE_w[A|\mathbb{B}]B) = w(BAB) \equiv w_A(B)$ for all $B \in \mathcal{P}_\mathbb{B}.$
\end{definition}

\begin{comment} According to the standard quantum mechanical formalism, after a measurement of $B \in \mathcal{P}_\mathbb{A}$, the state $W$ becomes $\hat{W} = BWB/w(B)$, hence the expected value of $A$ in this state is $w(A|B) = w(BAB)/w(B) = tr(BWBA)/tr(WB) = tr(\hat{W}A) = \hat{w}(A).$  This certainly is the quantum analogue of the comment made at the end of section (1.2)

Perhaps an interesting name for $\mathbb{B}$ is \textit{the measurement algebra}. This is the analogue of the classical $\sigma$-algebra $\sigma(Y)$ determined by the observation of a random variable $Y.$
\end{comment}

The result that allows us to think of conditional expectations as predictors is the following
\begin{theorem}\label{quantumprediction}
Assume that the sub algebra $\mathbb{B}$ is such that 
$$E_w[AC|\mathbb{B}] = E_w[A|\mathbb{B}]C\;\;\mbox{and}\;\; E_w[CA|\mathbb{B}] = CE_w[A|\mathbb{B}]$$
\noindent for any $C \in \mathbb{B},$ then $E_w[A|\mathbb{B}]$ is the best predictor of $A$ by an element of $\mathbb{B},$
\end{theorem}
\begin{comment}
It is easy to verify that when there exists a discrete  family $\{P_j° j \geq 1\}$ of projectors such that any $C \in \mathbb{B}$ can be written as $C = \sum c_jP_j,$ then the condition holds.
\end{comment}
\begin{proof}
We have to verify that$E_w[A|\mathbb{B}]$ is the minimizer of $E_w[(A - C)^2]$ when $C$ varies in $\mathbb{B}.$ The argument is the usual one, namely consider
$$E_w[\Big((A - E_w[A|\mathbb{B}]) + (E_w[A|\mathbb{B}] - C)\Big)^2]$$
\noindent expand the square, invoke the assumptions to get rid of the cross products and arrive at
$$E_w[(A - C)^2] = E_w[(A - E_w[A|\mathbb{B}])^2] + E_w[(E_w[A|\mathbb{B}] - C)^2]$$
\noindent which clearly achieves its minimum when $C = E_w[A|\mathbb{B}].$
\end{proof}

\section{Double slit like scenarios}

Consider, with no further specification, a system with an underlying Hilbert space $\mathcal{H}$ of dimension larger than 2, and suppose it is prepared in an initial state $W = |\psi_o><\psi_o|$ where $|\psi_o> = a|+> + b|->$ where $a$ and $b$ are complex numbers such that $|a|^2 + |b|^2 = 1.$
So, instead of a wave coming from infinity and impinging on a two holed screen, we create two waves emitted from two different points, but we do not know from which. By $|\pm>$ we denote respectively the wave (or particle) emitted from the upper, or respectively, the lower hole. Or think of an atom which may be in its ground state or in an excited state. Thus, the expected value of any observable $A$ with respect to the density operator $W$ (or in state $w$) is $tr(WA) = <\psi_o|A|\psi_o>.$

Let $B_+ = |+><+|$ and $B_- = |-><-|$ be the orthogonal projectors and assume that $|->$ and $|+>$ are orthogonal, and let $\mathbb{B}$ the (commutative) algebra generated by $B_+ $ and $B_-.$ 

According to theorem (\ref{quantumprediction}), the best predictor of an observable $A$ given the measurement algebra $\mathbb{B}$ is
\begin{equation}\label{predictor1}
E_w[A|\mathbb{B}] =  \frac{w(B_+AB_+)}{w(B_+)}B_+ + \frac{w(B_-AB_-)}{w(B_-)}B_-.
\end{equation}
We shall now consider some particular cases. Suppose that we consider $|+> = |\mathbf{x}_+> = |0,0,+1>$ and  $|-> = |\mathbf{x}_-> = |0,0,-1>,$ respectively the eigenstates of the position operator corresponding to a particle localized at either of $\pm 1$ along the $z$-axis, and we let $A = e^{-itH}|\mathbf{x}><\mathbf{x}|e^{itH}$. Here $H$ stands for the Hamiltonian of a particle of unit mass in units in which Planck's constant $\hbar$ is $1$.

Note to begin with that if we use the notation $K_t(\mathbf{x};\mathbf{x'}) = <\mathbf{x'}|e^{-itH}|\mathbf{x}>,$ then the expected value of $A$ in state $w$ is $w(A) = tr(WA) = |aK_t(\mathbf{x};|\mathbf{x}_+) + bK_t(\mathbf{x};\mathbf{x}_-)|^2$.
That is, if we do not observe which source is active, the probability of finding the particle at $x$  satisfies the standard "wave-particle duality" implied by the superposition principle.

If we decide to observe the position of the source, the result will be one of two possible values, and the predictor of $A$ given $\mathbb{B}$ is given in (\ref{predictor1}). To compute it explicitly note that
$$w(B_+AB_+) = tr(WB_+AB_+) = tr(B_+WB_+AB_+) = |aK_t(\mathbf{x};\mathbf{x}_+)|^2$$
\noindent and a similar looking expression is obtained for $w(B_-AB_-).$ Also, $w(B_+) = |a|^2,$ and so on. With all this, (\ref{predictor1}) is
\begin{equation}\label{predictor1.1}
E_w[A|\mathbb{B}]  = |K_t(\mathbf{x};\mathbf{x}_+)|^2B_+ + |K_t(\mathbf{x};\mathbf{x}_-)|^2B_-,
\end{equation}
\noindent and keep in mind that his is an observable. When we observe the position to be $(0,0,+1)$, then the initial state is reduced to $B_+WB_+/w(B_+)$ and the expected value of $E_w[A|\mathbb{B}]$ is just $|K_t(\mathbf{x};0,0,1)|^2$ as the standard analysis on the double slit experiment asserts. Observe also that the expected value of $E_w[A|\mathbb{B}]$ in the state $w$ is
$$w(E_w[A|\mathbb{B}]) = tr(WE_w[A|\mathbb{B}]) = |a|^2 |K_t(\mathbf{x};\mathbf{x}_+)|^2 + |b|^2|K_t(\mathbf{x};\mathbf{x}_-)|^2 $$
\noindent which is the classical expected value of the observed amplitude of particles emitted from $z = \pm 1$ with probabilities $|a|^2$ and $|b|^2$ respectively.

Had we insisted in seeing waves at $\mathbf{x}$, we should have considered $A = \frac{1}{E - H}|\mathbf{x}><\mathbf{x}|\frac{1}{E - H}.$ The analogue of (\ref{predictor1.1}) is
\begin{equation}\label{predictor1.2}
E_w[A|\mathbb{B}] = \phi(\mathbf{x} -\mathbf{x}_+)B_+ + \phi(\mathbf{x} -\mathbf{x}_-)B_-,
\end{equation}
\noindent where
$$\phi(\mathbf{x} -\mathbf{x}_+) = \frac{e^{i\omega||\mathbf{x} -\mathbf{x}_+||}}{ ||\mathbf{x} -\mathbf{x}_+||},$$
\noindent and $\omega = \sqrt{2E}.$ 

Again, if no observation is made as to where the source of particles is located, the expected value of $A$ in state $w$ is
$$w(A) = |a\phi(\mathbf{x} -\mathbf{x}_+) + b\phi(\mathbf{x} -\mathbf{x}_-)|^2,$$
\noindent whereas if we observe that the particle is at $\mathbf{x}_-$ then the observed signal at $\mathbf{x}$ is $\phi(\mathbf{x} -\mathbf{x}_-)$, that is the expected value of $E_w[A|\mathbb{B}]$ in the state $B_-WB_-/w(B_-).$ Or if we compute the expected value of $E_w[A|\mathbb{B}]$ in the state original state $w$, the result will be $|a|^2\phi(x-x_+) + |b|^2\phi(x-x_-)$ as classical physics would have predicted.

\section{Prediction in the presence of conservation laws}

Suppose now that our system is a composite systems, with Hilbert space $\mathcal{H} = \mathcal{H}_1\otimes\mathcal{H}_2$, and let $C$ be the selfadjoint operator on $\mathcal{H}$ denoting some conserved quantity, and suppose, to keep it simple that there is no interaction between the subsystems, and that $C = C_1 + C_2$ (or more properly, $C = C_1\otimes I_2 + I_1\otimes C_2$). We shall consider two variants of the same situation.

\subsection{Sch\"{o}dinger's cat type paradoxes} Suppose that the system is originally prepared in a state with density operate $W = |\psi_0><\psi_0|$ where this $\psi_0> = a|1,0> + b|0,1>$ where without fearing too much ambiguity, $C_1|0> = 0|0>$ and $C_1|1> = |1>$ and similarly for $C_2.$ Think about $|1,0><1,0|$ as describing the projector on a state with an excited atom and no photon, whereas $|0,1><0,1|$ describes projector on a state with the atom in its ground state plus a photon. In the second state the photon may be absorbed at a wall enclosing the system turning it blue(let the poor cat alone). If we look a the wall, and see it blue, the theory should predict that the atom has decayed. Again, the conditional expectation is not defined for all $A \in \mathbb{A},$ but the elements of the measurement algebra are easy to describe for the algebra is generated by two orthogonal projectors.

If we do not look at the wall (or do not make any observation about the photon) the prediction of the value of any observable $A$ in the state $W$ is $w(A) = tr(AW) =<\psi_0|A|\psi_0>.$ In particular, the probability of detecting a photon is $w(I_1\otimes |1><1|) = tr(WI_1\otimes |1><1|) = |b|^2$ which is the same as the probability of finding the atom in the ground state, namely $w(|0><0|\otimes I_2) = tr(W|0><0|\otimes I_2) = |b|^2.$ The observation algebra when we are looking at the photon is the algebra generated by the operators $B_0 = I_1\otimes |0><0|$ and $B_1 = I_1\otimes |1><1|.$ We do not consider more states of the photon number operator for the initial state that we prepared, there can be one photon at the most.

Note now that if a photon is observed, the computation to predict the probability of finding a photon is (with $\hat{W} = B_1WB_1/w(WB_1)$)
$$tr(E_w[w(|0><0|\otimes I_2\,|\,\mathbb{B}]B_1\hat{W}) = 1$$
\noindent as it should be, for if one photon is observed, the atom can only be in the ground state.

\subsection{EPR type paradoxes}
Our setup will be similar to the previous section. The change in notation for emphasis. That is, we shall consider a composite systems, with Hilbert space $\mathcal{H} = \mathcal{H}_1\otimes\mathcal{H}_2$, and let $L$ be the selfadjoint operator on $\mathcal{H}$ denoting some conserved quantity, and suppose, to keep it simple that there is no interaction between the subsystems, and suppose that $L = L_1 + L_2.$ Let us call $L_i$ the "spin" of the $i$-th particle and $L$ the total spin.

As initial state we consider $W = |\psi_o><\psi_o|$ where $|\psi_o> = a|+1,-1> + b|-1,+1>$ and assume that $L_i|\pm> = \pm |\pm>$ for $ i = 1,2.$ and to avoid issues related to degeneracy, assume that these eigenvectors are non-degenerate. Thus $L|\psi_o> = 0|\psi_o>.$ The given superposition only reflects the possibility that there are two possible ways in which the composite system can have total value of $L$ equal to $0.$

Again, we assume that the observation algebra is generated by the observation of the state of one the particles. For example, let it be the algebra $\mathbb{B}$ be generated by the projectors $B_- = I_1\otimes |1><1|$ and $B_- = |-1><-1|.$

If we observe the second particle to be $-1,$ with what probability is the spin of the first particle $+1$? The initial state reduces to $\hat{W} = B_-WB_-/w(B_-),$ and the pending computation is
$$[E_{\hat{w}}[|1><1|\otimes I_2\,|\,\mathbb{B}] = tr(B_-W |1><1|\otimes B_-)/ tr(WB_-) = 1$$
\noindent that is, once we have observed the second particle to have spin $-1$, we can assert that with probability $1$ the second the second particle has spin $1$. We do not have to measure it for it is a certain event.

\section{Concluding remarks} I hope to have convinced the reader that, as far as the paradoxes analyzed here goes, the whole mystery in quantum mechanics lies in the superposition principle, which in the examples treated enters in the specification of the initial states. When the result of observation is taken into account by proper conditioning as in classical probability, the paradoxes are removed, but the mystery associated with the superposition principle is there to stay.

\end{document}